\documentclass[11pt]{article}
\newsavebox{\PSLASH}
\sbox{\PSLASH}{$p$\hspace{-1.8mm}/}

\input{amssym}
\begin{document}
\title{ SLE($\kappa,\rho$)and Boundary Coulomb Gas}
\author{S.Moghimi-Araghi\footnote{samanimi@sharif.edu} ,M. A. Rajabpour\footnote{e-mail: rajabpour@mehr.sharif.edu}
, S. Rouhani \footnote{e-mail: rouhani@ipm.ir} \\ \\
Department of Physics, Sharif University of Technology,\\ Tehran,
P. O. Box: 11365-9161, Iran} \maketitle

\begin{abstract}
 We consider the coulomb gas model on the upper half plane with
different boundary conditions, namely Drichlet, Neuman and mixed.
We related this model to SLE($\kappa,\rho$) theories. We derive a
set of conditions connecting the total charge of the coulomb gas,
the boundary charges, the parameters $\kappa$ and $\rho$. Also we
study a free fermion theory in presence of a boundary and show
with the same methods that it would lead to logarithmic boundary
changing operators.
\end{abstract}
\section{Introduction}
Conformal field theories have found many applications in
classification of phase transitions and critical phenomena in two
dimensions and other areas such as string theory. In particular
the minimal models introduced in \cite{BPZ} reveal many exact
solutions to various two dimensional phase transitions like Ising
model at critical point or three critical Ising model and so on.
These models were first considered on the whole plane, but as many
surface phenomena are very interesting to analyze, boundary
conformal field theory was soon developed \cite{Cardy84}.
Essentially it was shown that conformal field theory in the half
plane with proper boundary condition (BC) could be mapped to a
whole-plane conformal field theory, the price you have to pay is
to insert image fields in the other half plane. The idea was then
applied to different problems such as turbulence \cite{turb}.

On the other hand, recently a new method to investigate the so
called geometrical phase transitions has been developed, which
were previously described by conformal field theories. The new
method, Stochastic Lowener Evolution (SLE) \cite{Schramm} is a
probabilistic approach to study scaling behavior of geometrical
models. For a review see \cite{Rhode,Lawler,kada,kager}. SLE's can
be simply stated as conformally covariant processes, defined on
the upper half plane, which describe the evolution of random
domains, called SLE hulls. These random domains represent critical
clusters.

The idea of SLE  was first developed by Schramm \cite{Schramm}. He
showed that under assumption of conformal invariance, the scaling
limit of loop-erased random walks is SLE$_2$. (SLE's are
parameterized by a real number $k$, and are abbreviated as
SLE$_k$) Also he claimed without proof that SLE$_6$ is the scaling
limit of critical percolation. The claim was proved later on by
Smirnov \cite{Smir}. He showed that in the scaling limit of
percolation, conformal invariance exists and also using the new
technic proved Cardy's formula \cite{cardy1,cardy2}.

As in SLE we are dealing with conformal mappings and geometrical
phase transitions, one expects that there should be a relation
between SLE and conformal field theory. The existence of a
boundary is an essential point in SLE, so one would expect it
would be related to BCFT's. An explicit relationship between the
two was discovered by Bauer and Bernard \cite{bau2,bau3,bau4,bau5}
and afterwards by \cite{RF1,RF2,RF3}. They coupled SLE$_\kappa$ to
boundary conformal field theory with specific central charge
depending on the parameter $\kappa$. These CFT's live on the
complement of SLE hulls in the upper half plane In such
situations, boundary states emerge on the boundary conformal field
theory. The good point is that these states are zero modes of the
SLE$_\kappa$ evolution, that is they are conserved in mean. This
means that all components of these states are local martingales of
SLE$_\kappa$ and hence one is able to compute crossing
probabilities in purely algebraic terms. The key point of these
result is existence of null vectors, just an in the case of
minimal models. It turns out that the Verma module in the CFT part
should contain level 2 null state, though higher level null
vectors have been considered, too \cite{ras,ma}. The level two
null state ,$|h\rangle$,  obeys the equation
\begin{equation}\label{Null2}
\left( 2L_{-2} +\frac{\kappa}{2} L_{-1}^2\right) |h\rangle=0,
\end{equation}
where $L_{-k}$'s are the Virasoro operatores. This equation
naturally leads to second order differential equations on
correlators containing the null state which are the same ones
obtained from stochastic approach.

More recently a generalization of these theories has emerged: the
so called SLE($\kappa,\rho$) \cite{lsw,dub}. The new theory
describes random growing interfaces in a planar domain which have
markovian property and conformal invariancy. In fact
SLE($\kappa,\rho$) is the minimal way to generalize the original
SLE while keeping self-similarity. We will come back to the
properties of such theories in the third section.

Cardy \cite{cardy04} extended the correspondence between
SLE$_\kappa$ and CFT, to the case of SLE($\kappa,\rho$). In this
case the equation in the CFT part is a little bit different from
equation (\ref{Null2}), it has an extra term which comes from a
current density $\mathcal{J}$:
\begin{equation}\label{ModNull}
\left( 2L_{-2} +\frac{\kappa}{2}
L_{-1}^2-\mathcal{J}_{-1}L_{-1}\right) |h\rangle=0.
\end{equation}
If a current density is already present in the theory, then the
equation would have a more physical meaning. The example
considered in \cite{cardy04} is a free bosonic field with the
action $(g/4\pi)\int (\partial \phi)^2 d^2 z$. The boundary
conditions are piecewise constant Drichlet; that is on real axis
we have
\begin{equation}\label{BoundDri}
\phi(x) = 2\pi \sum_j \alpha_j H(x-x_j)
\end{equation}
where $H(x)$ is the Heavyside step function.

The effect of a jump of strength $\alpha$ could be replaced with a
boundary condition changing operator $\phi_\alpha$. The state
produced by operation of this field is the one which satisfies the
equation (\ref{ModNull}). The correlation functions associated
with such field can be defined in the following way: first find a
classical solution satisfying the boundary condition (BC). Calling
the solution $\phi_{cl}$, we write $\phi= \phi_{cl}+\phi'$, where
$\phi'$ is assumed to vanish on the boundary. The partition
function is then calculated to be $Z=Z_{\alpha}Z'$. Note that this
happens, because the theory is free. In boundary conformal field
theory (BCFT) one can think of $Z_\alpha$ as the correlation
function $\langle \prod_j \phi_{\alpha_j}\rangle$.

Let's turn back to the specific problem. The classical solution of
equation of motion satisfying the boundary condition
(\ref{BoundDri} is
\begin{equation}
\phi_{cl}=i\sum \alpha_{j} \ln((z-x_{j})/(\bar{z}-x_{j})).
\end{equation}
Hence the partition function and the correlators are obtained to
be
\begin{equation}\label{CorrCar}
Z_\alpha=\left\langle \prod_j
\phi_{\alpha_j}\right\rangle=\prod_{j<k}\left(\frac{x_{k}-x_{j}}{a}\right)^{2g
 \alpha_{j}\alpha_{k}}.
\end{equation}
In a similar way the expectation value of energy-momentum density
in presence of such BC is found, which can be thought of the
correlation of energy-momentum tensor with the boundary fields
$\phi_\alpha$'s. Conformal Ward identity comes afterward with
which Cardy has found the action $L_0$, $L_{-2}$ and $L_{-1}^2$ on
the boundary fields. The fields are found to be primary with
scaling dimension is $h_i=g\alpha_i^2$. Going back to the
correlation functions (\ref{CorrCar}), one finds that the two
results are compatible if one assumes that the sum of all jump
vanishes.

To connect all these to SLE, one should look if it is possible to
satisfy either equation (\ref{Null2}) or (\ref{ModNull}). The
first one is satisfied if we assume $\kappa=4$ and
$\alpha_i^2=\alpha^{*\,2}=1/4g$, so that $h_i=1/4$. The other one
which is related to SLE($\kappa,\rho$) appears if we loose the
second condition, which leads to
\begin{equation}\label{SLE(kr)}
2L_{-2}\phi_{\alpha_i} = 2
L_{-1}^2\phi_{\alpha_i}-\sum_j{}^{\prime}\frac{\rho_j}{x_j-x_i}L_{-1}\phi_{\alpha_i},
\end{equation}
where $\rho_j= (\alpha_j-\alpha_i)(1-(\alpha_i/\alpha^*)^2)$ and
prime means that $j$ is excluded from the summation. This equation
can be written as (\ref{ModNull}) in terms of $J = -2i g^{1/2}
\partial \phi$ which is conserved due to equations of motion.

Cardy then mentions that it is possible to take Neuman boundary
condition (NBC) instead of Drichlet boundary condition (DBC), one
just have to consider the  dual description of the problem with
NBC on dual fields $\tilde{\phi}$ with insertions of vertex
operators $e^{i\alpha\tilde{\phi}}$.

In the next section, we will consider the coulomb gas action with
a charge at infinity and investigate different boundary
conditions. The next chapter is devoted to the connections to SLE
and SLE$(\kappa,\rho)$. In the last section we will apply the same
method to the fermionic action of $c=-2$ and find logarithmic
operators and some structures which has been found before
\cite{MRSalgeb}.

\section{Boundary Operators in Coulomb Gas Model}
Coulomb gas is defined via the following action:
\begin{equation}\label{coulomb action}
S=\frac{1}{4\pi}\int\left[g(\partial\varphi)^{2}+2\,Q\,R\varphi\right]
\end{equation}
where $R$ is the scaler curvature associated with the background
metric. We are interested in the case where $\varphi$ is defined
on the upper half plane. The boundary conditions can have several
forms: Dirichlet (DBC), Neuman (NBC) or mixed boundary condition,
but not all of them are conformally invariant.  A boundary
condition is conformally invariant if the antiholomorphic part of
the energy momentum tensor in the lower half plane is the analytic
continuation of its holomorphic part in the upper half plane. This
is satisfied if $T_{xy}=0$ at $y=0$ where
\begin{equation}\label{Txy}
T_{xy}=\frac{g}{2}\partial_{x}\varphi
\partial_{y}\varphi-\frac{ Q}{2}\partial_{x}\partial_{y}\varphi.
\end{equation}
In NBC case, where $\partial_{y}\varphi|_{y=0}=0$, the above
condition is automatically satisfied, while in DBC case, it is not
satisfied unless if $Q=0$. Let's take Neuman boundary condition,
we construct our desired boundary condition in the following way:
suppose we have some electrical charges at points $x_j$ located at
the boundary. We would like to find the boundary changing
operators associated with these electric charges. The classical
solution which satisfies Laplace equation together with the above
property has the form
\begin{equation}\label{neuman}
\varphi_{cl}=\sum \frac{\lambda_{j}}{g}
\ln(z-x_{j})(\bar{z}-x_{j})
\end{equation}
In this solution the normal derivative of field vanishes and hence
the solution satisfies NBC. For consistency, $\varphi$ has to
behave like $-2\, Q\: {\rm ln}(z \bar{z})$ when
$z,\bar{z}\rightarrow\infty$, so one should assume the condition
$\sum \lambda_{j}=-2 Q g$.

the next step is to calculate the partition function. If we write
$\varphi=\varphi_{cl}+\varphi^{\prime}$, with the condition
$\partial_{y}\varphi'|_{y=0}=0$, we'll have
$S[\varphi]=S[\varphi_{cl}]+S[\varphi']$, and the partition
function associated with $S[\varphi_{cl}]$ is derived to be
\begin{equation}\label{partition}
Z_{cl}=\prod_{j<k}\left(\frac{x_{k}-x_{j}}{a}\right)^{\frac {2
\lambda_{j}\lambda_{k}} {g}}.
\end{equation}
Here $a$ is the UV cut off. Let $Z'$ be the partition function
associated with $S[\varphi']$, then we define the boundary
correlation function to be
\begin{equation}\label{corrdef}
\langle\prod_{j}\phi_{\lambda_{j}}(x_{j})\rangle =\frac{Z}{Z'}
\end{equation}

Also any expectation value $\langle O\rangle$ in presence of this
boundary condition can be defined in the following way:
\begin{equation}\label{correlation}
\langle O\rangle =\frac{\langle
O\prod_{j}\phi_{\lambda_{j}}(x_{j})\rangle
}{\langle\prod_{j}\phi_{\lambda_{j}}(x_{j})\rangle}
\end{equation}
In particular the expectation value of energy-momentum tensor $T$
can be computed to give:
\begin{equation}\label{T}
\langle T\rangle =-g
(\partial\varphi_{cl})^{2}+2Q\partial^{2}\varphi_{cl}= \frac 1
g\sum_{j,k}\frac{\lambda_{j}\lambda_{k}}{(z-x_{j})(z-x_{k})}+2\frac
Q g \sum_{j}\frac{\lambda_{j}}{(z-x_{j})^{2}}.
\end{equation}
Taking the limit of $z\rightarrow x_i$ one can obtain the
conformal weights of the operators $\varphi(x_i)$,
$h_{\lambda_i}=\lambda_{i}(\lambda_{i}+2Q)/g$. The result suggests
that these operators are related to vertex operators
$\exp(i\lambda \phi)$ in the coulomb gas model.

Now let's investigate the Dirichlet boundary condition with $Q=0$.
The field is taken to be piecewise constant with some jumps $2\pi
\alpha$ at the points $x_{j}$'s. In this case, the classical
solution of the equation of motion is
\begin{equation}\label{Dirichlet}
\varphi_{cl}=i\sum\alpha_{j}\ln(\frac{z-x_{j}}{\bar{z}-x_{j}})
\end{equation}
Note that in the presence of external charge, second term of
offdiagonal energy-momentum tensor (\ref{Txy}) does not vanish at
the points $x_j$.

At this level, there are no conditions on $\alpha_{j}$'s, and
$Z_{cl}$ can be obtained as in the case of NBC. The result is just
the same as equation (\ref{partition}), you just need to transform
$g\rightarrow 1/g$ and $\lambda_j\rightarrow \alpha_j$. Similar to
the NBC case, there appear some operators on the points $x_{j}$.
Also the definition of correlation functions of such fields is the
same as equation (\ref{corrdef}). It can be shown that these
operators are related to magnetic vertices.

Again one can define the expectation values of any operator via a
relation similar to equation (\ref{correlation}). The related
fields are then found to be primary ones with conformal weight
$h_{\alpha_j}=g\alpha_{j}^{2}$. Now if we look back to the
correlation functions of $\varphi_{\alpha_i}$'s, one observes that
they are consistent with the weight derived, if the sum of all
jumps add up to zero, that is the value of $\varphi$ on both far
ends of the boundary should be the same\cite{cardy04}.

The other BC is the mixed one (with $Q=0$). In this case
$\varphi_{cl}$ has the following form:
\begin{equation}\label{Dirichlet2}
\varphi_{cl}=\sum \frac{\lambda_{j}}{g}
\ln(\bar{z}-x_{j})(z-x_{j})+i\sum\alpha_{j}\ln\left(\frac{z-y_{j}}{\bar{z}-y_{j}}\right)
\end{equation}
The above BC means that some electric and magnetic charges are
fixed on points $x_{j}$ and $y_{j}$. It turns out that in the
classical action, the two parts related to magnetic and electric
charges decouple. The results are a combination of the previous
two cases. If $x_{j}=y_{j}$ there is a electric-magnetic vertex
operator $O_{e,m}$ with the scaling dimension
$h=(\sqrt{g}\alpha+\lambda/\sqrt{g})^{2}$.  It can be argued that
for consistency we must have $\sum\lambda_{j}\alpha_{j}=0$ in
addition to previous conditions.

As in SLE, one can use similar bulk operators, rather than the
boundary ones, in brief we will talk about bulk correlation
functions. In the case of DBC one is able to use methods similar
to the image method in ordinary electromagnetic theory, so if
there is a magnetic charge at $z_j$ then another magnetic charge
with opposite sign should be located at $z^{*}_j$. Thus the
neutrality condition is satisfied automatically. But in Neuman
case, the image charge has the same sign and then the neutrality
condition is not satisfied automatically. The two point functions
of Dirichlet and Neuman BC's have the following form
\begin{eqnarray}\label{DN}
\langle O_{e} O_{-e}\rangle &=&\left(\frac{{\rm Im} z \,\,{\rm Im} w }{|z-w|^{2}|z-w^{*}|^2}\right)^{e^{2}} \\
\langle O_{m_{1}}O_{m_{2}}\rangle &=&\frac{1}{({\rm Im}
z)^{m_{1}^{2}}({\rm Im}
w)^{m_{2}^{2}}}\left|\frac{z-w}{z-w^{*}}\right|^{2m_{1}m_{2}}
\end{eqnarray}
These correlation functions could be used to develop a
SLE($\kappa,\rho$) theory.

\section{Relation to SLE($\kappa,\rho)$ }
In this section we would like to relate the result derived in the
previous section to the new and interesting topic
SLE(${\kappa,\rho}$). As we mentioned in introduction,
SLE($\kappa,\rho$) is somehow a more general case of
SLE$_{\kappa}$ which describes random growing interfaces in a
planar domain which have markovian property and conformal
invariancy.  This evolution has the following form
\begin{eqnarray}\label{sle}
\partial_{t}g_{t}(z)=\frac{2}{g_{t}(z)-W_{t}}, \hspace{0.5cm
}g_{0}(z)=z, \hspace{1cm}
dW_{t}=\sqrt{\kappa}dB_{t}+\sum_{1}^{n}\frac{\rho_{j}}{W_{t}-g_{t}(x_{j})},
\end{eqnarray}
where $x_j$'s are some arbitrary points not necessarily on the
boundary, and $g_{t}(z)$ is a conformal map from a subset of upper
half plane, $H_t$, to the whole upper half plane, $H$. In other
words, $H_t=H\backslash K \rightarrow H$ where $K_t$ is the
complement of $H_t$ in the upper half plane. The subset $K_{t}$
which is called the hull of evolution, is just the growing domains
mentioned above. Setting all of the parameters $\rho_{j}$ to zero,
one arrives at the ordinary SLE$_\kappa$, but with nonzero
$\rho_j$ we may have several random curves growing from the points
on the boundary \cite{BauBer05}. SLE($\kappa,\rho$) has many
interesting properties as you can see in \cite{lsw,dub}.

As we said earlier, Cardy \cite{cardy04} has established a
connection between SLE($\kappa,\rho$) and a CFT with Drichlet
boundary changing operators inserted at points $x_{j}$. He derives
the properties of the boundary operators $\phi_j$ which leads to
equation (\ref{SLE(kr)}) which is a specific case of the more
general equation
\begin{eqnarray}\label{nullsle}
\left(2L_{-2}-\frac{\kappa}{2}L_{-1}^{2}+\sum\frac{\rho_{j}L_{-1}}
{g_{t}(x_{j})}\right)\phi_{j}=0
\end{eqnarray}
needed to establish a connection to SLE($\kappa,\rho$).

Let's see how the different choices for boundary conditions in the
previous section affect the derived SLE($\kappa,\rho$). Taking
$g=1$, for the case of NBC, calculation of
$2L_{-2}-\frac{\kappa}{2}L_{-1}^{2}$ yields
\begin{eqnarray}\label{nullcoulomb}
2L_{-2}\phi_j-\frac{\kappa}{2}L_{-1}^{2}\phi_j=\hspace{10cm}\nonumber\\
\sum\frac{4\lambda_{i}(Q-\lambda_{j})+\kappa
\lambda_{i}\lambda_{j}}{(x_{i}-x_{j})^{2}}\phi_j+
(2-2\kappa\lambda_{i}^{2})\sum_{j,k}'\frac{\lambda_{j}\lambda_{k}}{(x_{i}-x_{j})(x_{i}-x_{k})}\phi_j
\end{eqnarray}
If  we impose the two conditions
$\displaystyle{\lambda_{i}^{2}=\frac{1}{\kappa}}=\lambda^2$ and
$Q=\displaystyle{\frac{\lambda(-\kappa+4)}{4}}$, then
$2L_{-2}-\frac{\kappa}{2}L_{-1}^{2}$ will be zero. Also we will
have
\begin{equation}
c=\frac{(6-\kappa)(3\kappa-8)}{2\kappa},\hspace{2cm}h=\frac{6-\kappa}{2\kappa}
\end{equation}
which is similar to the case of SLE$_{\kappa}$.

On the other hand, if we impose only the condition
$Q=\displaystyle{\frac{\lambda(-\kappa+4)}{4}}$ we will have
\begin{eqnarray}\label{nullcoulombn}
2L_{-2}\phi-\frac{\kappa}{2}L_{-1}^{2}\phi=(2-2\kappa\lambda_{i}^{2})
\sum_{j,k}'\frac{\lambda_{j}\lambda_{k}}{(x_{i}-x_{j})}L_{-1}\phi
\end{eqnarray}
which is the equation in SLE($\kappa,\rho$).

There is an equivalent to study this problem. Bauer, Bernard and
Kytola (BBK) have shown that SLE($\kappa,\rho$) could be obtained
starting from correlation functions of boundary fields
\cite{BauBer05}. Later Kytola has applied this method to the
coulomb gas model, that is he has considered the boundary
correlation functions of a coulomb gas and derived the
corresponding SLE($\kappa,\rho$)\cite{Kyt05}. This method,
although is a little bit different from the cardy's one, has the
same consequences. By means of our results in the previous
section, and using the method of BBK, one is able to derive the
corresponding SLE($\kappa,\rho$). Namely the correlations
(\ref{partition}) and the one obtained in other BC's are very
suitable to do this. Additionally, Schramm and Wilson
\cite{SchWil05} and also Friedrich and Bauer\cite{FriBau05},
argued that it not necessary to take the operators just on the
boundary and one is able to construct
SLE($\kappa,\rho_{bound.},\rho_{bulk}$) using both the operators
on the boundary and in the bulk. To do this, the correlation
functions (\ref{DN}) are helpful. The result are very similar to
the ones we have derived, you should just replace $x_j$ with $z_j$
where $z_j$'s are points in the upper half plane.

Another point to mention is that coulomb gas action is not the
only one to produce these results. Zamolodchikov and Zamolodchikov
\cite{ZZ} have considered the Liouville action
\begin{equation}\label{Liou}
S_{L}= \int \left(\frac{1}{4\pi}(\partial \phi)^2 + \mu e^{2 b
\phi}\right)
\end{equation}
with the boundary conditions
\begin{eqnarray}\label{BCLio}
\phi(z,\bar{z})&=& -2 \log z\bar{z} + O(1) \hspace{3.48cm}
|z|\rightarrow \infty \\
\phi(z,\bar{z})&=& -2 \eta_i \log (z-x_i)(\bar{z}-x_i) + O(1)
\hspace{1cm} |z|\rightarrow x_i
\end{eqnarray}
and have calculated the classical action to be
\begin{equation}
S_{L-cl}=\sum_{i<j} 2\eta_i\eta_j \log|x_i-x_j|^2
\end{equation}
where the condition $\sum \eta_i=1$ should be imposed. As you see
this is very similar to the correlation functions in the case of
coulomb gas and will lead to the same SLE$(\kappa,\rho)$.

Also Fateev, Zamolodchikov and Zamolodchikov \cite{FZZ} have
considered Liouville action on the disc with proper boundary
conditions and derived the correlation functions of the operators
in the bulk and on the boundary. Hence it should be related to a
radial SLE($\kappa,\rho$).

\section{Fermionic Boundary Changing Operators and LCFT's}
In this section we will apply the same method not to the action of
a bosonic field, but to the action of a free fermionic fields:
\begin{equation}\label{f-action}
S=\int \bar{\partial} \bar{\theta}\partial \theta
\end{equation}
where $\theta$ and $\bar{\theta}$ are grassman variables. This
action is related to $c=-2$ conformal field theory which is known
to be a logarithmic conformal field theory, that is, there can be
found logarithmic terms in correlation functions of the fields
inside the theory \cite{Gur}. In such theories there exists pair
of primary fields which transform into each other under conformal
transformations.

Before considering the action (\ref{f-action}) we would like to
review  a powerful method developed to investigate LCFT's, namely
the nilpotent weight method. The method is primarily based on a
composite field, containing the two primary fields and a nilpotent
variable which acts as a part of filed's weight\cite{MRSNil}, that
is, from the primary fields $\phi$ and $\psi$ we construct the
field $\Phi(z,\beta)=\phi(z)+\beta\,\psi(z)$ where $\beta$ is a
nilpotent variable. This fields acts as a primary field under
conformal transformations with the weight $h+\beta$, assuming
$\phi$ has weight $h$. The idea was then generalized in
\cite{MRSalgeb}, where the nilpotent variable was taken to be
product of a grassman variable and its conjugate, hence the
composite field contained four different fields, two bosonic and
two fermionic. This structure was observed before in $c=-2$ theory
by Kausch \cite{Kausch}. In this language the composite is written
to be
\begin{equation}\label{4field}
\Phi(z,\eta)=\phi(z)+\bar{\eta}\xi(z)+\bar{\xi}(z)\eta+\bar{\eta}\eta\psi(z)
\end{equation}
The fields $\phi$ and $\psi$ are the same as previous ones and
$\xi$ and $\bar{\xi}$ are the new primary fermionic fields. The
two point correlation function of such fields can be found
exploiting conformal invariance:
\begin{equation}\label{4FCor}
\langle \Phi(z_1,\eta_1)\Phi(z_2,\eta_2) \rangle
=\frac{(\bar{\eta}_1 +
\bar{\eta_2})(\eta_1+\eta_2)}{(z_1-z_2)^{2h+\bar{\eta}_1\eta_1+\bar{\eta}_2\eta_2}}
\end{equation}
This equation could be read through expanding both sides of it, in
terms of grassman variables to find correlation function of
individual fields.

With this brief review, we will go back to our problem, a $c=-2$
theory with different boundary conditions. Assume that the
boundary condition is Drichlet with discontinuities $\alpha_j$ at
points $x_j$. We should do the same procedure as in the bosonic
case. The classical solution satisfying this BC is
\begin{equation}\label{fersol}
\theta_{cl}=i\sum\alpha_{j}\ln\left(\frac{z-x_{j}}{\bar{z}-x_{j}}\right)
\end{equation}
Here $\alpha_j$'s are grassman variables in contrast with the
bosonic case. The next step is to calculate the classical part of
the partition function. As the solution (\ref{fersol}) is just the
same as the bosonic one, the result is the same, too. The only
difference is that we have $\bar{\theta}$ in the action and this
variable's jump is $\bar{\alpha}$ rather than $\alpha$. So we have
\begin{equation}\label{Zfercl}
Z_{\alpha}=\prod_{j\neq
k}\left(\frac{x_{k}-x_{j}}{a}\right)^{\bar{\alpha}_{j}\alpha_{k}}.
\end{equation}
Now we can assign this partition function to correlation of
boundary fields. Before doing this we have to impose a condition
similar to charge neutrality in the bosonic case, that is we
should assume that the sum of all the jumps has to be zero. To
assure that this happens we define
\begin{equation}\label{cor-def}
\langle \prod_j \vartheta(\alpha_j,x_j)\rangle=Z_{\alpha}\times
\delta(\sum_j \alpha_j)\delta(\sum_j \bar{\alpha}_j).
\end{equation}
Here $\vartheta_j$'s are the boundary fields and could be expanded
in terms of $\alpha_j$'s to have $\vartheta(x,\alpha)= \phi(x)
+\bar{\alpha} \xi (x) + \bar{\xi}(x)\alpha
+\bar{\alpha}\alpha\psi(z)$. Let's investigate the two point
correlation of such fields. As the delta function of a grassman
variable is itself, we'll have
\begin{equation}\label{2point}
\langle \vartheta
(x_1,\alpha_1)\vartheta(x_2,\alpha_2)\rangle\propto
\left({x_{1}-x_2}\right)^{(\bar{\alpha}_{1}\alpha_{2}+\bar{\alpha}_{2}\alpha_{1})}
(\bar{\alpha}_1+\bar{\alpha}_2)(\alpha_1+\alpha_2)
\end{equation}
Comparing with equation (\ref{4FCor}) and noting that sum of
$\alpha_j$'s is vanishing, we observe that we have found the
correct result. Note that the weight of this field is zero as
expected in $c=-2$ theory. It may be possible to construct a non
zero weight boundary field if we let the action (\ref{f-action})
have a bosinic part. Higher correlation functions could be also
derived in the same way, but now we would like to find the OPE of
the fields with energy-momentum tensor. The expectation value of
energy momentum tensor could be read easily through
\begin{equation}\label{fT}
\langle T(z) \rangle_{\alpha} =
-\bar{\partial}\bar{\theta}_{cl}\partial\theta_{cl}=
\sum_{j,k}\frac{\bar{\alpha}_j{\alpha_k}}{(x_j-x_k)}.
\end{equation}
which leads to the OPE:
\begin{equation}
T(z)\vartheta(x_j,\alpha_j)=\frac{\bar{\alpha}_j{\alpha_j}}{(z-x_j)^2}
+\frac{2}{z-x_j}\sum{}'
\left(\frac{\bar{\alpha}_{j}\alpha_{k}}{x_j-x_k}+
\frac{\bar{\alpha}_{k}\alpha_{j}}{x_k-x_j}\right)+\ldots ,
\end{equation}
which means that $\vartheta$ fields have weight equal to
$\bar{\alpha}\alpha$, which is consistent with previous results
\cite{MRSNil,MRSalgeb}. Using these boundary operators and their
correlation functions, one is able to find the related
SLE$(\kappa,\rho)$, exploiting the method introduced by Bernard
{\it et al.} \cite{BauBer05}.

\end{document}